\documentclass[aps,prl,groupedaddress,twocolumn]{revtex4-1}
\usepackage{natbib}
\usepackage{graphicx}
\usepackage{color}
\usepackage{bm}
\usepackage{longtable}
\usepackage{amsmath} 
\usepackage{amsfonts}
\usepackage{amssymb}
\usepackage{hyperref,enumerate}
\newcommand{\Tr}{\text{Tr}}

\begin{document}
\setlength{\abovedisplayskip}{3.5pt}
\setlength{\belowdisplayskip}{3.5pt}

\title{Unsharp measurements, joint measurability and classical distributions for some qudits}


\author{H S Smitha Rao$^1$}
\email[]{smitharao@ycm.uni-mysore.ac.in}
\author{Swarnamala Sirsi$^1$ }
\author{Karthik Bharath$^2$}
\affiliation{$^1$ Yuvaraja's college, University of Mysore, Mysuru, India }
\affiliation{$^2$ University of Nottingham, Nottingham, U.K.}


\begin{abstract}
Classicality associated with joint measurability of operators manifests through a valid classical joint probability distribution on measurement outcomes. For qudits in dimension $n$, where $n$ is prime or power of prime, we present a method to construct unsharp versions of projective measurement operators which results in a geometric description of the set of quantum states for which the operators engender a classical joint probability distribution, and are jointly measurable. Specifically, within the setting of a generalised Bloch sphere in $n^2-1$ dimensions, we establish that the constructed operators are jointly measurable for states given by a family of concentric spheres inscribed within a regular polyhedron, which represents states that lead to classical probability distributions. Our construction establishes a novel perspective on links between joint measurability and optimal measurement strategies associated with Mutually Unbiased Bases (MUBs), and formulates a necessary condition for the long-standing open problem of existence of MUBs in dimension $n=6$.
\end{abstract}


\maketitle
\textbf{Introduction}.
Joint measurability \cite{Busch09, Heinossari08} of Positive Operator-Valued Measures (POVMs), which generalise projective measurements, has been shown to characterise classicality, through its connections to quantum steering \cite{roope14, uola15}, incompatibility\cite{Chen}, non-locality\cite{Brunner14}, contextuality\cite{ghosh}, no-signalling theory\cite{wolf09}.  A complementary notion of classicality associated with measurement operators, more adherent to traditions of classical probability theory, is related to quantum characteristic functions\cite{scully1, wigne,MHill} from which classical probability distributions can be derived.  Some aspects of the relationship between joint measurability and classical probability distributions have been explored \cite{akr, UR08, Busch13}. Pertinently, commuting operators are jointly measurable, and lead to joint probability distributions via characteristic functions. Throughout, for brevity, we use joint distribution to refer to a valid classical joint probability distribution on a set of outcomes.

On the other hand, projective operators constructed using Mutually Unbiased Bases (MUBs) are considered to be maximally incompatible \cite{schwinger} and nonclassical, prohibiting simultaneous measurements. However, using an unsharpness parameter $\eta$, jointly measurable POVMs, known as unsharp measurements,  based on projectors from eigenbases of the Pauli matrices were constructed in \cite{Busch87,Busch09} for two-level systems or qubits.  Relationships between joint measurability and MUBs were examined in \cite{Uola16,Brunner19,Heino12}, and have recently been exploited to quantify incompatibility \cite{DSFB} and state discrimination \cite{CHT}. 

An explicit construction of jointly measurable POVMs based on MUBs for general $n$-level systems, or qudits, would offer key insights into the interplay between classicality and nonclassicality of the corresponding operators, and their influence on (joint) probabilities associated with measurement outcomes. In this paper, we show how this can be achieved for qudits in dimension $n$ where $n$ is a prime or power of prime; existence of MUBs are known only for such dimensions. Armed with an orthonormal matrix basis obtained from MUBs\cite{SRao19}, in a manner similar to the qubit case, we construct unsharp measurements using a parameter $\eta$ that are POVMs and jointly measurable; this results in a joint distribution on measurement outcomes. We then use the same orthonormal basis to define a valid quantum characteristic function, derive the corresponding joint distribution, and show that it is related to the joint distribution arising from the POVMs quite simply through the unsharpness parameter $\eta$.  

What do the constructed operators based on an unsharpness parameter $\eta$ imply about the quantum states, since after all, measurement is with respect to a fixed state? To answer this, we consider $n$-dimensional density matrices $\rho(\vec \theta)$ parameterised  by an $N=(n^2-1)$-dimensional vector $\vec \theta$. Within the geometric setting of a generalised Bloch sphere in dimension $N$ representing all states, we identify two regions: a family of concentric spheres with radii $\eta\leq N^{-1/2}$ that corresponds to states with respect to which the constructed operators are jointly measurable;  and, a regular polyhedron with vertices on surface of the Bloch sphere, containing the preceding family of spheres, which corresponds to states that engender joint distributions obtained through characteristic functions. An important consequence of our geometric description is that it provides a necessary condition for the existence of MUBs in composite dimensions, which is a long-standing open problem.

\emph{Joint measurability}. A POVM is a collection $\mathbb{E}=\{E(x)\}$ of self-adjoint operators, generalising projective operators, that satisfy $E(x) \geq 0$ and $\sum_x E(x)=\mathbb{I}$, where $x$ denotes an outcome. With respect to $E(x)$, an outcome $x$ occurs with probability $\Tr[\rho E(x)]$. Joint measurability can be defined for a collection of POVMs; however, for our purposes it suffices to consider the definition for a single one: for a set of outcomes $\lambda:= \{x_1,\ldots,x_k\}$, the operators $\mathbb{E}=\{E(x_1),\ldots,E(x_k)\}$ are jointly measurable if and only if there exists a global POVM $\mathbb{G}=\{G(\lambda): 0 \leq G(\lambda) \leq \mathbb{I}; \sum_\lambda G(\lambda)=\mathbb{I}\}$, and positive constants $\{a_\lambda(x_i)\}$ \cite{ACHT} such that:
\[
E(x_i)=\sum_{\lambda} a_\lambda(x_i) G(\lambda); \quad \sum_{\lambda} a_\lambda(x_i)=1.
\]
It was shown in \cite{roope14} that failure of joint measurability of $\mathbb{E}$ implies nonclassicality. It's important to note that joint measurability depends on a fixed quantum state $\rho$: it may happen that $\mathbb{E}$ be jointly measurable with respect to $\rho_1$ but not with respect to another quantum state $\rho_2$. 

\emph{Quantum characteristic functions}.  
For a $k$-dimensional classical random variable $\vec Y$ with joint distribution $p(\vec y)=p(y_1,\ldots,y_k)$, the Fourier transform 
\[\phi(\vec t)=\int e^{i\vec{t}\cdot\vec{y}}p(\vec {y})\text{d} \vec y \quad \text{or} \quad
\phi(\vec t)=\sum_{\vec y}e^{i\vec{t}\cdot\vec{y}}p(\vec {y})
\]
is referred to as its characteristic function, depending on whether $p(\vec y)$ is a continuous or discrete distribution. By virtue of its definition, $\phi$ uniquely determines $p$ through the inverse Fourier transform. 

If we view a vector  $\vec X=(X_1,\ldots,X_k)$ of operators $X_k$ as a quantum analogue of a classical random vector, noncommutativity implies that there are multiple ways to  define $e^{\vec t\cdot\vec X}$, and hence the characteristic function $\phi$. This problem is typically addressed using symmeterisation rules, popular amongst which are the Margenau-Hill \cite{MHill} rule, which, for example when $k=3$, proposes
\begin{align*}
\label{MM}
e^{i\vec t\cdot\vec X} \longrightarrow
 \frac{1}{3!} \sum_{\pi \in \Pi_3} \Big[e^{it_{\pi(1)}{X_{\pi(1)}}}
 e^{it_{\pi(2)}{X_{\pi(2)}}}e^{it_{\pi(3)}{X_{\pi(3)}}}\Big],
\end{align*}
where $\Pi_k$ is the symmetric group of permutations of $\{1,\ldots,k\}$ with bijections $\pi: \{1,\ldots,k\} \to \{1,\ldots,k\}$, and the Wigner-Weyl \cite{wigne} rule, which proposes $e^{i\vec t\cdot\vec X} \to e^{i\vec t\cdot\vec X}$ for any fixed chosen ordering of $X_1,\ldots,X_k$. The Margenau-Hill rule results in a discrete joint distribution (probability mass function) while using the Wigner-Weyl rule results in a  continuous joint distribution (probability density function) \cite{GR}. For a chosen symmetrisation rule the quantum characteristic function associated with a state $\rho$ and operators $\vec X$ is then defined as $$\phi(\vec t)=\Tr [\rho e^{i\vec t\cdot\vec X} ].$$

Irrespective of the symmetrisation rule, the crucial aspect of such a definition of $\phi$ is that the map $ \vec t \mapsto \phi(\vec t)$, unlike the situation with classical random variables, is not guaranteed to be the Fourier transform of a valid probability distribution $p(\vec x)$ on $\mathbb{R}^k$ for every fixed state $\rho(\vec \theta)$; this a consequence of Bochner's theorem \footnote{Bochner's theorem: $\phi(\vec t)$ is a valid characteristic function if and only if for every $r$-tuple $(\vec t_1,\ldots, \vec t_r)$ the $r \times r$ matrix with entries $\phi(\vec t_i -\vec t_j), i,j=1,\dots,r$ is non-negative definite and Hermitian}\footnote{See Example 4.1 in \cite{partha} and discussion therein.}. \\

\textbf{Geometric perspective for qubits}.  
We first consider the qubit case, for which the corresponding density matrix assumes the form
$\rho(\vec \theta)=\frac{1}{2}(\mathbb{I}_{2}+ \vec{\sigma} \cdot \vec \theta)$ where $\vec \sigma=(\sigma_x,\sigma_y,\sigma_z)$ with $\sigma_{i}$, $i=1, 2, 3$ denoting the  well-known Pauli operators, and the components of Bloch vector $\vec{\theta}$ are such that $\theta_{i}= Tr(\rho \sigma_i)$. The constraint $Tr[\rho(\vec \theta)^{2}] \leq 1$ implies that $\theta_{1}^{2} + \theta_{2}^{2} + \theta_{3}^{2}  \leq 1 $, with equality attained only for pure states. Thus the set of density matrices for qubits can be identified with the famous Bloch sphere $S^2(\vec \theta):=\{\vec \theta: \theta_{1}^{2} + \theta_{2}^{2} + \theta_{3}^{2}  \leq 1 \}$ with the surface of the sphere corresponding to pure states.

\emph{Joint measurability}. 
The eigenbases of the Pauli operators are MUBs, and result in optimal measurements \cite{pawel, woot}. Jointly measurable unsharp operators can be constructed from the Pauli operators based on the spectral decomposition 
\begin{equation*}
\sigma_{i}= \sum_{x_{i} \in \{-1,1\}}x_{i} \hat{P}(x_{i}), \quad i=1,2,3,
\end{equation*} 
where $x_{i}$ are the eigenvalues of $\sigma_{i}$ and $\hat{P}(x_i)$  are the projection operators corresponding to the eigenstates of $\sigma_{i}$. Thus, each projection operator is of the form, 
\begin{equation*} 
\label{p1}
{\hat{ P}} (x_{i})= \frac{1}{2}(\mathbb{I}_{2}+x_{i}\sigma_{i}), \quad i=1, 2, 3,
\end{equation*}
with associated probabilities 
$$p(x_{i})= \Tr[\rho(\vec \theta) {\hat{ P}} (x_{i})]= \frac{1}{2}(1+x_{i}\theta_{i}).$$
This in turn implies that  $-1 \leq \theta_{i} \leq 1$. The form of each projection operator implies that $\theta_i, i=1,2,3$ cannot be obtained simultaneously. 
Joint measurement of these parameters is possible by introducing a noise parameter and considering operators 
\begin{equation*}
E(x_{i})= \frac{1}{2}(\mathbb{I}_{2} + \eta x_{i} \sigma_{i}), \quad i= 1, 2, 3, 
\end{equation*}
where $0 \leq \eta \leq 1$ is referred to as the unsharpness parameter; when $\eta=1$ we recover the projection operators $\hat{P}(x_i)$.  It was shown in \cite{Busch86} that the operators $\mathbb{E}:=\{E(x_1), E(x_2),E(x_3)\}$ are jointly measurable with global POVM $\mathbb{G}=\{G(x_1,x_2,x_3): x_i=\pm 1,i=1,2,3\}$,
\begin{equation}
\label{g3}
G(x_1,x_2,x_3)= \frac{1}{8}\left[\mathbb{I}_{2}+ \eta (x_1\sigma_{1} + x_2\sigma_{2} +x_3\sigma_{3} )\right],
\end{equation}
only when $\eta \in (0,1/\sqrt{3}]$. For a state $\rho(\vec \theta)$,  the corresponding joint probability distribution associated with the jointly measurable operators $\mathbb{E}$ is thus
\begin{align}
\label{g4}
p(x_1,x_2,x_3)&= \Tr [\rho(\vec \theta) G(x_1,x_2,x_3)] \nonumber\\
&= \frac{1}{8}\left[1+ \eta (x_1 \theta_1+ x_2\theta_2 +x_3\theta_3)\right]. 
\end{align}
From the constraint $\theta^2_1+\theta^2_2+\theta^2_3 \leq 1$ on the Bloch vector $\vec \theta$, we thus see that the set of states $\rho(\vec \theta)$ with respect to which $\mathbb{E}$ is jointly measurable can be identified with a family of spheres $\{S^2(\eta \vec \theta), \eta \in (0,1/\sqrt{3}]\}$ within the Bloch sphere $S^2(\vec \theta)$ sharing the same origin. 


 \emph{Joint distribution}.
Motivated by the discrete spectrum of the Pauli operators $\sigma_i, i=1,2,3$, we consider the Margenau-Hill symmetrisation rule in order to define a quantum mechanical characteristic function $\phi(\vec t)$ with $\vec t=(t_1,t_2,t_3)$ for every state $\rho(\vec \theta)$ on the Bloch sphere; we then seek  a classical probability distribution $p(x_1,x_2,x_3)$ that corresponds to the inverse Fourier transform of $\phi(\vec t)$. It's worth noting that this approach is quite different to ones that seek a spin quasidistributions for classical random variables taking values on $S^2$, as considered in \cite{scully2, GR}. 
Using the Margenau-Hill symmetrisation rule on Pauli operators, it was shown in \cite{Sirsi07} that the function
\begin{equation}
\label{charac}
\phi (t_{1}, t_{2}, t_{3})= \frac{1}{3!} \Tr[\rho (\beta_{123}+\beta_{132}
+\beta_{213} +\beta_{231} +\beta_{312} +\beta_{321})],
\end{equation} 
 where $\beta_{abc}= e^{it_{a}\sigma_{a}}e^{it_{b}\sigma_{b}}e^{it_{c}\sigma_{c}}$ with $a,b,c=\{1,2,3\}$,
is the characteristic function of classical random variables $(X_1,X_2,X_3)$ with joint distribution 
 \[
 p(x_1,x_2,x_3)=\frac{1}{8}(1+x_{1}\theta_1+x_{2}\theta_2+x_{3}\theta_3),
 \]
 where $x_i \in \{-1,1\}, i=1,2,3$, which coincides with the distribution in \eqref{g4} for $\eta=1$. The Bloch vector $\vec \theta$ additionally satisfies $|\theta_1|+|\theta_2|+|\theta_3| \leq 1$. 
 Evidently then the relevant region $\mathcal O(\vec \theta)$ is an octahedron within the Bloch sphere with the six vertices $(\pm 1, 0, 0)$, $(0, \pm 1, 0)$, $(0, 0, \pm 1)$ on surface of the sphere corresponding to pure qubit states, associated with the eigenvalues of $\sigma_{1}, \sigma_{2}, \sigma_{3}$.  Summarily, for each  $\vec \theta \in \mathcal{O}(\vec \theta)$, the classical characteristic function associated with $(X_1,X_2,X_3)$ will coincide with the quantum characteristic function $\phi(\vec t)$. 
 
\begin{figure}
	\centering
	\includegraphics[width=0.24\textwidth]{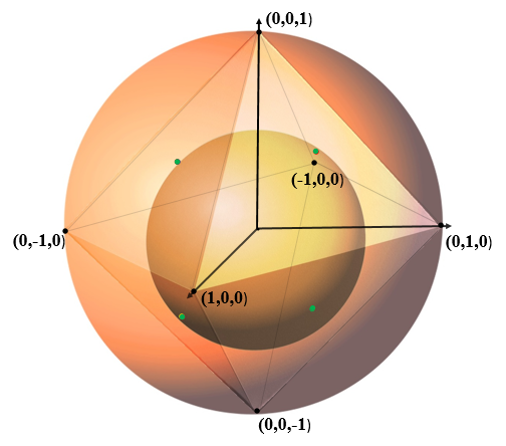}
	\caption{\small{Within the Bloch sphere $S^2(\vec \theta)$, the Octahedron (black vertices) contains states for which valid joint probability distributions on measurement outcomes can be constructed; the insphere $S^2((1/\sqrt{3})\vec \theta)$ of the octahedron contains states for which jointly measurable operators $\mathbb{E}$ can be constructed. 
	The insphere is tangent to the equi-triangular faces at eight centroids, four of which are shown in green}.}
		\label{pic}
\end{figure}

Figure \ref{pic} provides a graphical representation of the combined geometric description. We see that the unsharp measurements $\mathbb{E}$ cannot be jointly measurable with respect to any pure state on the surface of the Bloch sphere. On the other hand, from the octahedron $\mathcal{O}(\vec \theta)$ we see that every  state with a valid joint distribution can be represented as a convex combination of the 6 pure states corresponding to its vertices; notably this includes the set of states corresponding to the family on spheres of radii $\eta \leq 1/\sqrt{3}$. 
The sphere $S^2((1/\sqrt{3}) \vec \theta)$ associated with joint measurability of $\mathbb{E}$, is the insphere of the octahedron, tangent to each of the 8 two-dimensional faces (equilateral triangles with sides of length $\sqrt{2}$) at their respective centroids $c_i(\vec \theta) \in \mathcal{O}(\vec \theta)$ for $i=1,\ldots,8$. The corresponding states $\rho(c_i(\vec \theta))$ represent the only states on the lower-dimensional faces with respect to which the operators in $\mathbb{E}$ are jointly measurable. 
The radius $1/\sqrt{3}$ is the maximum value of the unsharp parameter $\eta$ under which the $E_i$ are jointly measurable.  \\


\textbf{Geometric perspective for qutrits}.
The upshot to the detailed description given for qubits is that exposition of geometric picture for qutrits is simplified. Crucial to the construction for qubits is the fact that eigenstates of Pauli operators are MUBs. 
For qutrits, and higher-level systems, using arbitrary $\mathfrak{su}(n)$ Lie algebra generators do not necessarily lead to analogous unsharp measurements $\mathbb{E}$. 
For example, use of the Gell-Mann orthonormal basis set $\{\hat{\Lambda}_i, i=1,\ldots,8\}$ results in a global POVM
	\begin{equation*}
	G(\lambda_1, \lambda_{2}, \ldots, \lambda_8)=\frac{1}{3^{7}}\Big[\mathbb{I}_{3}+\eta(\lambda_{1} \hat{\Lambda}_{1}+\lambda_{2} \hat{\Lambda}_{2}+ \ldots +\lambda_{8} \hat{\Lambda}_{8})\Big].
	\end{equation*}
   where $\{\lambda_{i}\}$ are the set of eigenvalues associated with $\Lambda_i$, each set containing three eigenvalues. However, it is not possible to represent the marginal POVM $E(\lambda_i)$ as unsharp versions (using a single parameter $\eta$) of the projectors associated with the matrices    \footnote{E.g: For $\hat{\Lambda}_1$, $E({\lambda}_1)= \sum_{\lambda_2,..,\lambda_8 }G(\lambda_{1},\lambda_{2},\ldots, \lambda_8)=(\mathbb{I}_3+\eta\lambda_1 \hat{\Lambda}_1)/3$. But $E({\lambda}_1)$ doesn't reduce to the case of projection operator associated with $\Lambda_1$ when $\eta=1$.}. This implies that the corresponding geometric picture of a family of spheres within the Bloch sphere is unavailable; moreover, the construction of the corresponding regular polyhedron linked to joint probability distribution is patently opaque. We instead consider the MUB-driven operators $\{\hat \alpha_i, i=1,\ldots,8\}$ proposed in \cite{SRao19}: 
	
{\footnotesize
	\begin{align*}
	\hat{{\alpha}}_{1}&=  {\sqrt{\frac{3}{2}}} \left(\begin{array}{ccc}
	1 & 0 & 0\\
	0 & 0 & 0\\
	0 & 0 & -1\\
	\end{array} \right),
	\qquad
	\hat{\alpha}_{2}=  \frac{1}{\sqrt 2} \left(\begin{array}{ccc} 
	1 & 0 & 0\\
	0 & -2 & 0\\
	0 & 0 & 1\\
	\end{array} \right),\\
	\hat{\alpha}_{3}&= {\frac{1}{\sqrt{2}}} \left(\begin{array}{ccc} 
	0 & -i\omega & i\omega^{2}\\
	i\omega^{2} & 0 & -i\omega\\
	-i\omega & i\omega^{2} & 0\\
	\end{array} \right), 
	\quad
	\hat{\alpha}_{4}= \frac{1}{\sqrt 2}\left(\begin{array}{ccc} 
	0 & -\omega & -\omega^{2}\\
	-\omega^{2} & 0 & -\omega\\
	-\omega & -\omega^{2} & 0\\
	\end{array} \right),\\ 
	\hat{\alpha}_{5}&={\frac{1}{\sqrt{2}}} \left(\begin{array}{ccc} 
	0 & -i  & i\omega^{2} \\
	i  & 0 & -i  \omega^{2} \\
	-i \omega & i \omega & 0\\
	\end{array} \right), 
	\quad
	\hat{\alpha}_{6}=\frac{1}{\sqrt 2} \left(\begin{array}{ccc} 
	0 & -1 & -\omega^{2}\\
	-1 & 0 & -\omega^{2} \\
	-\omega & -\omega & 0\\
	\end{array} \right),\\
	\hat{\alpha}_{7}& ={\frac{1}{\sqrt{2}}} \left(\begin{array}{ccc} 
	0 & -i \omega^{2} & i\omega^{2} \\
	i \omega & 0 & -i\\
	-i \omega & i  & 0\\
	\end{array} \right),
	\quad
	\hat{\alpha}_{8}= {\frac{1}{\sqrt{2}}} \left(\begin{array}{ccc} 
	0 & -\omega^{2} & -\omega^{2}\\
	-\omega & 0 & -1\\
	-\omega & -1 & 0\\
	\end{array} \right).
	\end{align*}
}

Since $\Tr(\hat{\alpha}_{i} \hat{\alpha}_{j})= 3 \delta_{i, j}$, the $\{\hat \alpha_i\}$  form an orthonormal basis for Hermitian matrices; they also form a mutually disjoint, maximally commuting set since they can be divided into four sets of two, within which the two operators commute: $(\hat \alpha_i, \hat \alpha_{i+1})$ commute for $i=1,3,5,7$. They are Pauli-like in the sense that their eigenbases are MUBs; we refer to \cite{SRao19} for details.  The density matrix $\rho(\vec \theta)$ can expanded in terms of the $\hat{\alpha}_i$ as
\[
\rho(\vec \theta)= \frac{1}{3}[\mathbb{I}_{3} + \sum_{i= 1}^{8} \theta_{i} \hat{\alpha}_{i}],
\] where $\theta_{i}= \Tr[\rho(\vec \theta) \hat{\alpha}_{i}]$.  The condition $\Tr [\rho(\vec \theta)^2] \leq 1$ implies that $\sum_{i=1}^8 \theta^2_i \leq 2$, and a similar Bloch-like seven-dimensional sphere $S^7(\vec \theta)$ of radius $\sqrt{2}$ in eight dimensions emerges. Bounds on the parameters are given by $-\sqrt{\frac{3}{2}} \leq \theta_{i} \leq \sqrt{\frac{3}{2}} $ when $i= 1, 3, 5, 7$, and $-{\sqrt{2}} \leq \theta_{j} \leq \frac{1}{\sqrt{2}}$ when  $j= 2, 4, 6, 8$. Since the $\hat \alpha_i$ comprise of four sets of two commuting operators, we have three pairs of eigenvalues as measurement outcomes
\[
\left\{\left(\sqrt{\frac{3}{2}}, \sqrt{\frac{1}{2}}\right), \left(0, -\frac{2}{\sqrt{2}}\right),
	\left(-\sqrt{\frac{3}{2}}, \sqrt{\frac{1}{2}}\right)\right\}
	\]
shared between the operators, denoted as $z=(x,y), z'=(x',y'), z''=(x'',y'')$ and $z'''=(x''',y''')$, where
\begin{align*}
x&= x' =x''=x'''=\sqrt{\frac{3}{2}}, 0, -\sqrt{\frac{3}{2}}\thinspace ;\\
y&= y'= y''=y'''=\sqrt{\frac{1}{2}}, -\frac{2}{\sqrt{2}}, \sqrt{\frac{1}{2}}\thinspace.
\end{align*}


\emph{Joint measurability}.
Since $\{\hat{\alpha}_i\}$ are based on MUBs, and are maximally commuting, we are left to consider only four unique measurement outcomes. We can construct unsharp measurements 
$\mathbb{E}:=\{E(z),E(z'),E(z''),E(z''') \}$ in an identical manner to the qubits using projection operators
\begin{align*}
\hat{P}(z)&= \frac{1}{3}(\mathbb{I}_{3}+x \hat{\alpha}_{1} + y \hat{\alpha}_{2}); \\
\hat{P}(z')&= \frac{1}{3}(\mathbb{I}_{3}+x' \hat{\alpha}_{3} + y' \hat{\alpha}_{4});\\
\hat{P}(z'')&= \frac{1}{3}(\mathbb{I}_{3}+x'' \hat{\alpha}_{5} + y'' \hat{\alpha}_{6});\\
\hat{P}(z''')&= \frac{1}{3}(\mathbb{I}_{3}+x''' \hat{\alpha}_{7} + y'''\hat{\alpha}_{8}).
\end{align*}
Arguments along identical lines as with the qubits confirm that $\mathbb{E}$ are jointly measurable 
with respect to a  global POVM $\mathbb{G}= \{G(z,z',z'',z''')\}$ with
\begin{align}
\label{g5}
   G(z,z',z'',z''')&= \frac{1}{81} \Big[\mathbb{I}_{3} + \eta\big(x \hat{\alpha}_{1} + y \hat{\alpha}_{2}+ x'\hat{\alpha}_{3} + y' \hat{\alpha}_{4} \nonumber\\
   &+ x'' \hat{\alpha}_{5} + y'' \hat{\alpha}_{6}+ x'''\hat{\alpha}_{7} + y''' \hat{\alpha}_{8} \big)], 
\end{align}
 only when $\eta \in (0,1/\sqrt{8}]$.
The geometric picture for joint measurability is hence again a family of spheres $S^7(\eta \vec \theta)$ within the Bloch sphere of radii $\eta \leq 1/\sqrt{8}$.

\emph{Joint distribution}.
The task here is to define a quantum characteristic function $\phi(\vec t)$ using the operators $\{\hat \alpha_i, i=1,\ldots,8\}$ such that its Fourier inversion results in a valid classical joint probability mass function on eight classical random variables $\vec X=(X_1,\ldots,X_4,Y_1,\ldots,Y_4)$, wherein $X_i$ assume values in $\{x, x',x'',x'''\}$ and the $Y_i$ take values in $\{y,y',y'',y'''\}$. We again employ the Margenau-Hill symmetrisation rule. The key advantage in using the $\{\hat{\alpha}_i\}$, in contrast to Gell-Mann matrices for instance, is that the four sets of two commuting operators can be combined into four operators for the purpose of defining a quantum characteristic function. For a fixed $\vec t =(t_1,\ldots,t_8)$, let 
\begin{align*}
\hat{A}_{1}&= \hat{\alpha}_{1} t_{1}+\hat{\alpha}_{2}t_{2},\quad  \hat{A_{2}}= \hat{\alpha}_{3} t_{3}+\hat{\alpha}_{4}t_{4}\\
\hat{A}_{3}&= \hat{\alpha}_{5} t_{5}+\hat{\alpha}_{6}t_{6}, \quad \hat{A}_{4}= \hat{\alpha}_{7} t_{7}+\hat{\alpha}_{8}t_{8}. 
\end{align*}
Note that $\hat{A}_{1}$ is diagonal and Hermitian. Since the transformation from $\hat{A}_i$ to $\hat{A}_j$ corresponds to a unitary transformation from an MUB to another, we see that $\hat{A}_{2}$, $ \hat{A}_{3}$ and $\hat{A}_{4}$ can also be diagonalised, respectively, with unitary transformations $\hat{U}_j=1/\sqrt{3}M_j, j=2,3,4$, where

 {\footnotesize
\begin{align*}
M_2=\left(
\begin{array}{ccc}
  1 & 1 & 1  \\
1 & \omega & \omega^{2} \\
1 & \omega^{2} & \omega \\
\end{array}\right),
M_3 =	 \left(
\begin{array}{ccc}
  1 & \omega^{2} & 1  \\
1 & 1 & \omega^{2} \\
1 & \omega & \omega \\
\end{array}\right),
M_4 =\left(
\begin{array}{ccc}
  1 & \omega & 1  \\
1 & \omega^{2} & \omega^{2} \\
1 & 1 & \omega \\
\end{array}\right), 
\end{align*}
}
with $\omega= e^{2\pi i/3}$. Conversion to diagonal form facilitates the definition of a quantum characteristic function using the technique  in \cite{Devi1994}, resulting in 
\begin{equation*}
    \phi(\vec t)= \frac{1}{4!}\sum _{\pi \in \Pi_4} \Tr\left[\rho(\vec \theta) \left(\beta(\pi(1)\pi(2)\pi(3)\pi(4))\right)\right],
\end{equation*}
where $\beta(abcd)= e^{i\hat{A}_{a}}e^{i\hat{A}_{b}}e^{i\hat{A}_{c}}e^{i\hat{A}_{d}}$.  Such a definition of $\phi(\vec t)$ ensures that its Fourier inverse
\begin{align*}
{\label{MH2}}
p(z,z',z'',z''')= \frac{1}{3^4}& \Big[1 + (x {\theta_{1}} + y{\theta_{2}}+ x'{\theta_{3}} + y' {\theta_{4}} \\
&+ x'' {\theta_{5}} + y''{\theta_{6}}+ x'''{\theta_{7}} + y'''{\theta_{8}} )\Big]
\end{align*}
is a valid distribution on the classical random vector $\vec X$; $p(z,z',z'',z''')$ again coincides with the distribution in \eqref{g5} when $\eta=1$. The form of the matrices $M_1, M_2$ and $M_3$, the constraint $\sum_{i=1}^{8}|\theta_{i}|^{2} \leq 2$, and the presence of four commuting pairs implies that the  $4 \times 3=12$ values of $\vec \theta$ on the surface of the Bloch sphere that represent pure states are given by $1/\sqrt{2}$ times the coordinates
\begin{align*}
&(\sqrt{3},1,\vec 0_6), 
(-\sqrt{3},1,\vec 0_6), (\vec 0_6, \sqrt{3},1),(\vec 0_4,-\sqrt{3},1,\vec 0_2),\\
&(\vec 0_6, -\sqrt{3},1),(0,-2,\vec 0_6), (\vec 0_3,-2,\vec 0_4), (\vec 0_5,-2,\vec 0_2), (\vec 0_7,-2),\\
&(\vec 0_2,\sqrt{3},1,\vec 0_4), (\vec 0_2,-\sqrt{3},1,\vec 0_4), (\vec 0_4,\sqrt{3},1,\vec 0_2),
\end{align*}
where $\vec 0_r$ denotes a vector of $r$ zeroes.
We observe that each vertex has two vertices with which it subtends an angle $2 \pi/3$ at the origin, and is orthogonal to the rest; the two vertices are the `diametrically opposite' points on the Bloch sphere linked by an $SU(3)$ rotation. Thus  each vertex is  formed by four mutually orthogonal equilateral triangular planes in the Bloch sphere. The regular polyhedron has 
$3^4=81$ faces and 54 edges, and represents the region within which $\phi(\vec t)$ matches the classical characteristic function obtained from $p(z,z',z'',z''')$. 

A combined geometric  picture similar in spirit to the qubit case arises. Joint measurability of the operators $\mathbb{E}$ is restricted to the set of states $\rho(\eta \vec \theta)$ with $\eta \in (0,1/\sqrt{8}]$ implying a geometric description of a family of spheres of radii $\eta$. The sphere with radius $1/\sqrt{8}$ is the insphere of the regular polyhedron $\mathcal{P}(\vec \theta)$, tangent to each of the 81 face at the centroids.  From such a geometric description we note that states with respect to which the unsharp $\mathbb{E}$ are jointly measurable engender valid joint probabilities on measurement outcomes. \\

\textbf{General qudits}.
The proposed geometric exploration of classical behaviour of quantum systems of dimension $n>3$ can be carried out along quite similar lines, \emph{as long as corresponding  MUBs are known to exist} \cite{SRao19}.  For example, MUBs exist for $n=4$ dimensions (power of prime \cite{woot}), and we can construct 15 operators $\{\hat{\alpha}_i\}$: five sets of three commuting operators. Following the above construction of unsharp measurements, within the Bloch sphere $S^{14}(\vec \theta)$, we obtain a similar picture of a family of spheres with radii upper bounded by $1/\sqrt{15}$ within a regular polyhedron with 20 vertices ($5 \times 4$ eigenvalue sets), 160 edges and $4^5=1024$ faces. The existence of complete set of MUBs for composite dimension systems is unknown. Several attempts at an answer for the lowest composite $n=6$ have concluded that there cannot exist more than three sets of MUBs \cite{zan,gra,62,61,64}. However, when viewed purely geometrically, a corresponding regular polyhedron within a Bloch sphere in $35$ dimensions should exist \cite{Beng05}; explicit construction of such a polyhedron using the methods proposed here would necessarily then imply existence of corresponding MUBs. \\

\textbf{Acknowledgements}. HSS thanks the Department of Science and Technology (DST), India for the grant of INSPIRE Fellowship. KB acknowledges partial support from grants NSF DMS 1613054 and NIH R01 CA214955.
\bibliography{joint}
\end{document}